\documentclass[hidelinks]{eptcs}
\usepackage{url}
\usepackage{graphicx}
\usepackage[utf8x]{inputenc} 
\usepackage{epstopdf}
\usepackage{mathtools}
\usepackage{amssymb}
\usepackage{amsthm} 
\usepackage{gensymb}
\usepackage{bm}
\usepackage{xspace}
\usepackage{stmaryrd}
\usepackage[shortlabels]{enumitem}
\usepackage{comment}
\usepackage[dvipsnames]{xcolor}
\usepackage{bussproofs}
\usepackage[tracking=true]{microtype}
\usepackage[paper,notion,hyperref,quotation]{knowledge}
\specialcomment{note} { \begin{tcolorbox} \begin{paragraph}{Comment:}} { \end{paragraph}\end{tcolorbox} }

\excludecomment{remov}
\excludecomment{note}
\usepackage{tcolorbox}

\usepackage{mdframed}
\usepackage{todonotes}
\usepackage{wrapfig}
\usepackage{csquotes}
\usepackage{letltxmacro}
\usepackage{ifthen}
\usepackage{xifthen}
\usepackage[titletoc]{appendix}
\usepackage[labelformat=simple]{subcaption}
\usepackage{scalerel}
\usepackage{etoolbox}

\newtoggle{extendedversion}
\togglefalse{extendedversion}
\newcommand{\extendedversion}[1]{\iftoggle{extendedversion}{#1}{}}

\title{A Tractable Logic for Molecular Biology}

\author{Adrien Husson\qquad\qquad Jean Krivine
  \institute{
Université de Paris, IRIF, CNRS\\
F-75013 Paris, France}
\email{\quad husson@irif.fr\quad\qquad jean.krivine@irif.fr}}
\def\titlerunning{A Tractable Logic for Molecular Biology}
\def\authorrunning{A. Husson \& J. Krivine}

\begin{document}
\newtheorem{theorem}{Theorem}[section]
\newtheorem{lemma}[theorem]{Lemma}
\newtheorem{prop}{Proposition}
\newtheorem{proposition}{Proposition}[section]
\newtheorem{definition}{Definition}[section]

\addtocontents{toc}{\protect\setcounter{tocdepth}{0}}

\setcounter{definition}{0}
\renewcommand{\thedefinition}{}

\newenvironment{example}
{\vspace{1em}\noindent\textit{Example}\\}
{\vspace{1em}}
\maketitle
\begin{abstract}
  We introduce a logic for knowledge representation and reasoning on protein-protein interactions. Modulo a theory, formulas describe protein structures and dynamic changes. They can be composed in order to add or remove static and dynamic observations. A second-order circumscription operator then enables nonmonotonic reasoning on the changes implied by a formula. We introduce deduction rules that produce formulas which are, up to equivalence, in a first-order fragment with decidable satisfiability and validity. Importantly, the rules can produce circumscribed formulas.
\end{abstract}



\newcommand{\SigmaTr}{\vv{\Sigma}}
\newcommand{\Tsbhf}{T^n_{\Sigma}}
\newcommand{\Tbhf}{T^n_{\SigmaTr}}
\newcommand{\Tfull}{\mathcal{T}^n_{\SigmaTr}}

\newcommand{\reffig}[1]{Figure~\ref{#1}}
\newcommand{\refsec}[1]{Section~\ref{#1}}

\newcommand{\mc}[1]{\mathcal{#1}}
\newcommand{\map}{\triangleright{}}

\newcommand{\FV}{\text{FV}}

\newcommand{\im}{\text{Im}}

\makeatletter
\newcommand\newsubcommand[2]{\newcommand#1[1]{##1\sc@sub{#2}}}
\def\sc@sub#1{\def\sc@thesub{#1}\@ifnextchar^{\sc@mergesubs}{^{\sc@thesub}}}
\def\sc@mergesubs^#1{^{\sc@thesub#1}}
\makeatother

\newsubcommand{\pt}{\star}

\newsubcommand{\mt}{(\star)}

\newcommand{\ball}{\mathcal B}
\newcommand{\adj}{\ell}
\newcommand{\ballA}{\ball_{\mdl A}}
\newcommand{\adjA}{\ell_{\mdl A}}

\newcommand{\set}[1]{\{#1\}}

\newcommand{\of}[2]{#1({\mdl{#2}})}

\newcommand{\form}[1]{\text{{\fontfamily{lmss}\selectfont #1}}}
\newcommand{\iform}[1]{\textit{#1}}
\newcommand{\ffform}[1]{\text{{\fontfamily{lmss}\selectfont #1}}}

\newcommand{\udf}{\lightning}

\providecommand{\mathbold}{\mathbf}
\newcommand{\mb}[1]{\mathbold{#1}}

\newcommand{\tuple}[1]{\mb{#1}}


\newcommand*{\dfn}{\mathrel{\vcenter{\baselineskip0.4ex \lineskiplimit0pt
                           \hbox{\scriptsize.}\hbox{\scriptsize.}}}%
                                                =}

\newcommand{\lnk}{\raisebox{3pt}{$\scriptstyle{\frown}$}} 
\newcommand{\dolnk}{\raisebox{3pt}{$\scriptstyle{\frown +}$}} 
\newcommand{\brk}{\mathrlap{\lnk}\ ^/}

\newcommand{\fa}[1]{\forall #1.\ }
\newcommand{\ex}[1]{\exists #1.\ }

\newcommand{\Du}{\blacktriangle}
\newcommand{\Dr}{\blacktriangleright}
\newcommand{\Dd}{\blacktriangledown}
\newcommand{\Dl}{\blacktriangleleft}

\newcommand*{\textdoubletrianglever}{%
  \resizebox{!}{\heightof{X}}{%
    \vbox{%
      \hbox{$\blacktriangledown$}%
      \nointerlineskip
      \kern.15ex
      \hbox{$\blacktriangle$}%
    }%
  }%
}

\newcommand*{\textdoubletrianglehor}{%
  \resizebox{!}{\heightof{X}}{%
    \vbox{%
      \kern.4ex
      \hbox{$\blacktriangleright\blacktriangleleft$}%
      \nointerlineskip
      \kern.4ex
    }
  }%
}

\newcommand{\changed}[1]{\mc C_{\overline{#1}}}

\newcommand{\rim}{\text{right}_{\mdl A}}
\newcommand{\upm}{\text{up}_{\mdl A}}

\newcommand{\sig}[1]{#1}
\newcommand{\mdl}[1]{\mathfrak{#1}}

\newcommand{\diff}[1]{\Delta\!#1}

\newcommand{\circum}[1]{\downarrow\!#1}
\newcommand{\kircum}[1]{\dagger\,#1}
\newcommand{\dom}{\text{dom}}

\newcommand{\vr}{\text{var}}
\newcommand{\rs}{\text{restr}}
\newcommand{\mn}{\text{min}}
\newcommand{\fx}{\text{fix}}

\newcommand{\fulltuple}
  {\tuple{P_\mn},\tuple{P_\rs},\tuple{f_\rs},
  \tuple{P_\vr},\tuple{f_\vr},\tuple{P_\fx}}

  \providecommand{\cite}{TODO}
  \providecommand{\citet}[2][]{TODO}
\providecommand{\citet}[2][]
{\citeauthor{#2}~\ifthenelse{\equal{#1}{}}{\shortcite{#2}}{\shortcite[#1]{#2}}}
\providecommand{\citeauthoryear}{}

\newcommand{\sem}[2]{\llbracket #1 \rrbracket_{#2}}
\newcommand{\semsimple}[1]{\llbracket #1 \rrbracket}
\newcommand{\trees}{\ell}

\newcommand\ScaleExists[1]{\vcenter{\hbox{\scalefont{#1}$\exists$}}}


\newcommand{\safe}{\mc V}
\newcommand{\asdist}{\rightsquigarrow}

\newcommand{\qm}{\odot}

\newcommand{\tu}{\ \text{\adjustbox{trim=0pt {.5\height} 0pt 0pt,clip,raise={.2\height}}{\ensuremath{\updownarrow}}}\!\,}
\newcommand{\td}{\ \text{\adjustbox{trim=0pt 0pt 0pt {.5\height},clip}{\ensuremath{\updownarrow}}}\!\,}
\newcommand{\tud}{\ \updownarrow\!\!}


\newcommand{\dedu}[1]{\textsc{\textls[-30]{#1}}}
  \newcommand{\varof}[1]{\langle#1\rangle}

  \newcommand{\addrem}[3]{(#1\cup#2)\!\setminus\!#3}
\newcommand\sbullet[1][.65]{\mathbin{\vcenter{\hbox{\scalebox{#1}{$\bullet$}}}}}
\newcommand\consep{\,;}

\newcommand{\actordeq}{\trianglelefteq}
\newcommand{\actord}{\vartriangleleft}
\newcommand{\circordeq}{\preceq}

\newcommand{\added}{\raisebox{.1em}{\footnotesize $\oplus$}}
\newcommand{\removed}{\raisebox{.1em}{\footnotesize $\ominus$}}


\newcommand{\Tn}{\mc T^n}
\newcommand{\Tns}{\mc T^n_{\text{\tiny supp}}}
\newcommand{\Tnfb}{\mc T^n_{\text{\tiny FB}}}

\newcommand{\exall}{\exists^*\mb{|}\forall^*}

\newcommand{\Dyn}{Dyn}
\renewcommand{\Dyn}{D\kern-0.05em y\kern-0.11em n}
\newcommand{\Stat}{S\kern-0.05em t\kern-0.05em a\kern-0.05em t}

\newcommand{\psiAFull}{\psi_1(\form{P}\cup \form{P'},\tuple{\Dyn'})}
\newcommand{\psiBFull}{\psi_2(\form{P}\cup \form{P'},\tuple{\Dyn'})}

\newcommand{\hiddensubsubsection}[1]{
  \addtocontents{toc}{\protect\setcounter{tocdepth}{2}}
  \subsubsection{#1}
  \addtocontents{toc}{\protect\setcounter{tocdepth}{3}}
}

\newcommand{\tocthm}{}

\newcommand{\toclemma}[1]{%
  \addtheocontentsline{Lemma \ref{#1} (\thelemma\ in appendix)}%
  \textit{Lemma \ref{#1} in paper}%
}

\newcommand{\toctheorem}[1]{%
  \addtheocontentsline{Theorem \ref{#1} (\thetheorem\ in appendix)}%
  \textit{Theorem \ref{#1} in paper}%
}

\newcommand{\actmin}{%
\hstretch{0.7}{\nabla\!}%
}

\knowledge{rule:forall}{text={\dedu{$\forall$-Guard}},notion}
\knowledge{rule:exists}{text={\dedu{$\exists$-Guard}},notion}
\knowledge{rule:weak}{text={\dedu{Weak}},notion}
\knowledge{rule:bool}{text={\dedu{Bool}},notion}
\knowledge{rule:inv}{text={\dedu{Invariant}},notion}
\knowledge{rule:dynamic}{text={\dedu{Dynamic}},notion}
\knowledge{rule:static}{text={\dedu{Static}},notion}
\knowledge{rule:circum}{text={\dedu{Circumscribe}},notion}

\section{Introduction}





Molecular biology accumulates data and mechanisms suspected to play key roles in the cellular ecosystem. The activity of discovery currently outpaces human abilities to follow and collate new mechanisms. For instance, p53, a protein family relevant to cell apoptosis and cancer formation, is mentioned in the title or in the abstract of about 4700 papers for the year 2018 alone (PubMed).

In 2014, DARPA financed a large program named ``Big Mechanism'', for about \$45M, pointing explicitly to the problem of extraction and integration of molecular biology facts from biological literature \cite{Cohen2015}. 
Along this line of research, 
our intention is to provide a formal basis for describing structural and dynamic biological knowledge suitable for composition and reasoning. To illustrate the type of knowledge we are aiming at, here is a typical sentence from a molecular biology paper:

\begin{displayquote}
  \emph{``The activation of Raf-1 by activated Src requires phosphorylation of Raf-1 on Y340 and/or Y341 [...].  Tyrosine phosphorylation and activation of Raf-1 have been shown to be coincident. However, others have been unable to detect phosphotyrosine in active Raf-1.''}. \cite{mason1999serine}
\end{displayquote}

At this level of abstraction, proteins are considered as chains of amino acid residues such as Y340 and Y341, which are identified by their type (Y for tYrosine) and their position in the chain (resp. 340 and 341). Proteins have names, here Raf-1 and Src, and are usually divided into domains  or regions that are covering sub-sequences of amino acids. Domains may also be given a name. For instance, Raf-1 has a ``Zinc finger'' domain in the 137-183 region.\footnote{%
uniprot.org/uniprot/P09560%
}
  
Importantly, \emph{static} names of proteins, domains and residues can be completed with \emph{dynamic} attributes. Here, ``phosphorylation'' denotes the attachment of a phosphate group to a protein residue, which tends to modify the protein structure. One then talks about a phosphorylated protein, a phosphorylated domain or, as in the example above, a phosphorylated residue. Other dynamic attributes such as ``active'' are commonplace in molecular biology.

Underlying the snippet of biological literature given above is the notion of protein interactions: ``the activation of Raf-1'' by ``activated Src'' indicates that Raf-1 and active Src can bind to each other so that phosphorylation of Raf-1 by Src may occur. Stable binding of proteins requires complementary domains that stick together with various affinities. The binding state of a protein (or a region) is therefore also a dynamic, relational property.

We express observations using formulas. Their models are \emph{transitions}, which represent a biological change from a precondition state to a postcondition state. 
States are forests of linked trees. Trees encode proteins.
The root of a tree represents the entire protein, and children represent sub-parts of the protein. Nodes can have static names (Raf-1, Src,  Y340, \ldots) and dynamic attributes  ($\form{Phos}$, $\form{Active}$, \ldots).

  The logic we introduce in this paper has the following design constraints:\\
  -- The logic describes changes in a compositional way and works as a basis for knowledge representation.\\
  -- One is in principle able to run queries on the information to judge the impact of adding new knowledge to an existing base.\\
  -- The logic accomodates both knowledge collation and biological \emph{modelling}, which applies a parsimony assumption on available biomechanisms. This corresponds to commonsense reasoning: changes not implied by observations cannot occur. This assumption is expressed with a second-order operator on formulas. The formalism introduced in this paper allows one to mix both activities, knowledge collation and modelling, in a single logic while maintaining queryability.

\paragraph{Overview}
We represent mixtures of proteins (states) as labelled forests. The trees have bounded height and degree. A root $x$ represents a whole protein, and children represent domains, subdomains or residues depending on their height in the tree. Transitions are pairs of forests, with overlapping underlying sets. They represent one step of biological change. The first element of a transition is the precondition, and the second element is the postcondition. Static labels (Src, Raf-1) are not allowed to change during the transition, and neither does the structure of the trees. Dynamic labels ($\form{Phos},\form{Active}$) may change. We encode changes by copying each dynamic predicate: for instance, $\form{Phos}(x)$ means ``$x$ is phosphorylated in the precondition'', while $\pt{\form{Phos}}(x)$ means ``$x$ is phosphorylated in the postcondition''.

The other dynamic aspect is a functional and symmetric relation $\form{Link}$ which represents protein-protein interactions, typically noncovalent bonds. Functionality captures the fact that binding sites are resources, so $\form{Link}(x,y)$ is incompatible with $\form{Link}(x,z)$ if $y\neq z$\@. It comes with a copy for the postcondition, $\pt{\form{Link}}(x,y)$\@. If $x$ represents a protein connected to multiple partners, the corresponding links are distributed on separate children nodes of $x$\@.

A transition contains zero or more changes (edge removal, label change, etc). Transitions can be ordered along their changes. Intuitively, if two transitions have the same precondition and one contains all the changes of the other, then they are comparable along a \emph{change order}. We introduce nonmonotonic reasoning with the operator $\downarrow$: for a formula $\phi$, $\circum\phi$ denotes the models $\phi$ that are minimal along the change order. In models of $\circum\phi$, no unnecessary changes occur.
 
\begin{example}
Consider the following formula:
\[
  \textit{Observation}_1(x) \dfn 
  \fa{y} \pt{\form{Link}}(x,y) \rightarrow \neg\pt{\form{Active}}(\textit{parent}(x)) 
\]

where $\textit{parent}(x)$ is a term denoting the parent of $x$ in its tree (or $x$ itself, in the case where $x$ is a root). The models of $\textit{Observation}_1(x)$ are all transitions such that if, in the postcondition, the interpretation of $x$ is $\form{Active}$, then $x$ is not linked to any protein (it is `free') also in the postcondition.
We can compose observations and for example add the observation that $x$ ends up connected to an Src protein:
\begin{align*}
  \textit{Observation}_2(x) &\dfn
  \ex{z} \text{Src}(z) \wedge \pt{\form{Link}}(x,z)\\
  \textit{Obs}(x) &\dfn \textit{Observation}_1(x) \wedge \textit{Observation}_2(x)
\end{align*}
%
\begin{figure}
  \caption{Minimal ($\mdl T$) and non-minimal ($\mdl T'$) transitions for $\iform{Obs}$}
  \begin{subfigure}[b]{0.49\textwidth}
    \includegraphics[width=\textwidth]{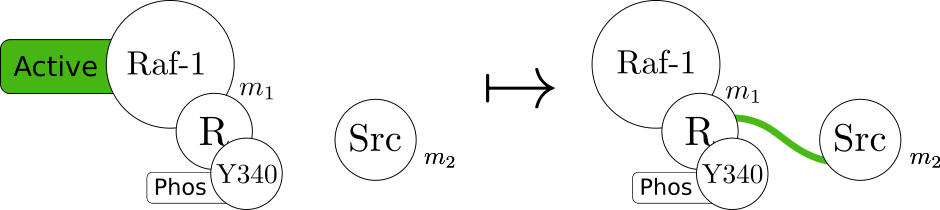}
    \caption{$\mdl T \vDash \textit{Obs}(m_1)$ and $\mdl T \vDash \circum\textit{Obs}(m_1)$}
    \label{f:simple_illustration}
  \end{subfigure}%
  \hfill%
  \begin{subfigure}[b]{0.49\textwidth}
    \includegraphics[width=\textwidth]{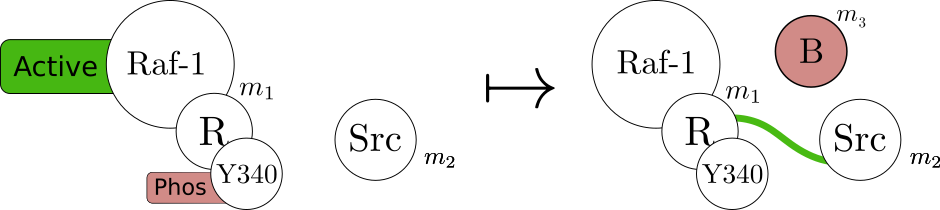}
    \caption{$\mdl T' \vDash \textit{Obs}(m_1)$ and $\mdl T' \nvDash \circum\textit{Obs}(m_1)$}
    \label{f:iota_example_bad}
  \end{subfigure}
\end{figure}%
%
%
\hspace{1em}In figure \ref{f:simple_illustration}, we show the transition $\mdl T$, which satisfies $\textit{Obs}(m_1)$\@. The precondition is on the left of the arrow, and the postcondition is on the right. Changes are highlighted in green. $m_1$ is the region R bearing the Y340 residue of Raf-1, which is phosphorylated. The Raf-1 node, $\textit{parent}(m_1)$, loses the label $\form{Active}$\@. A link between $m_1$ and $m_2$ is created.

There is, so to speak, an open-world assumption on changes: for instance, a transition that has thousands of trees in the precondition and deletes them all in the postcondition satisfies $\textit{Observation}_1$\@.
While knowledge collation has to be made with an open world assumption (more structure or more changes might be added when more knowledge is accessible), modelling focuses on dynamics and is an activity that is intrinsically parsimonious with regards to changes: the dynamics of a model are restricted to what is implied by current knowledge. Therefore, we also want the ability to reason with the additional assumption that \emph{all relevant observations have been made}. To remove transitions with spurious changes, we use the operator $\downarrow$.
Transitions that are models of $\circum \textit{Obs}(x)$ have the changes required for $x$ to be linked to an Src in the postcondition, and such that either $x$ is free or $x$'s parent is inactive in the postcondition --- and no other changes.

In figure \ref{f:iota_example_bad}, we see the transition $\mdl T'$, which satisfies $\textit{Obs}(m_1)$\@. $\mdl T$ and $\mdl T'$ have the same precondition, but $\mdl T'$ contains additional changes, highlighted in red: the tyrosine residue of $m_1$ becomes unphosphorylated, and a new protein $m_3$ is created. Intuitively, those additional changes are not required by $\textit{Obs}$, and we will show that $\mdl T'$ does not satisfy $\circum\textit{Obs}(m_1)$.
\end{example}
%

\paragraph{Structure of the paper}
Section \ref{s:preliminaries} introduces the vocabulary. Section \ref{s:transitions_flbs} describes two classes of structures, \emph{transitions} and \emph{transitions of forests of linked bounded trees}, or \emph{FLBs}. 
Modulo the theory of FLBs, first-order satisfiability is not decidable, but satisfiability in the $\exists^*\forall^*$ prenex fragment is.
Section \ref{s:change_minimisation} defines a \emph{change order} $\actordeq$ on transitions; two transitions with the same precondition are comparable if all the changes of one are included in the other.
We denote the change-minimal models of a formula $\phi$ with $\circum\phi$\@.
%
Section \ref{s:deduction_rules} introduces deduction rules and states the main theorem: modulo the theory of \emph{supported FLBs}, deducible formulas are in a class with decidable satisfiability and validity. 
Section \ref{s:proof_sketch} defines the new constructs of \emph{unified circumscription} and \emph{preservation}, which allow one to prove the main theorem. Both of general relevance, the former can express $\circum\phi$ as a second-order formula, and the latter, through model-theoretic properties, defines a class of $\phi$ for which $\circum\phi$ is actually first-order.

\section{Preliminaries}\label{s:preliminaries}
Formulas are first-order except when otherwise noted. Equality is allowed, but not constant symbols. Signatures are noted $\Sigma, \Theta,\mathellipsis$, structures are noted, $\mdl A, \mdl M, \mathellipsis$, and interpretations from a set of variables to a structure domain are noted $\mu, \nu, \mathellipsis$\@.

By ``nodes'' we mean elements in the domain of a structure.

  For $\pi$ a first-order quantifier prefix and $\phi$ any formula, $\phi \in \pi$ to mean that $\phi$ is equivalent to a prenex first-order formula with quantifier prefix in $\pi$\@. 

If $\vartheta$ is a term, a set of terms, or a formula, $\varof \vartheta$ is the set of variables mentioned in $\vartheta$\@.
For a structure $\mdl A$, a term $t$, and $\mu:\varof t \rightarrow \dom(\mdl A)$, $\sem{t}{\mdl A,\mu}$ is the interpretation of $t$ in $\mdl A$ under $\mu$. We also lift the semantic brackets to sets of terms. Moreover, for $\phi(\tuple x)$ a formula over the tuple of variables $\tuple x$, $\sem{\phi(\tuple x)}{\mdl A}$ is the set of tuples $\tuple a$ from $\dom(\mdl A)$ such that $\mdl A \models \phi(\tuple a)$\@.


\paragraph{Transitions} Transitions are a generic framework for representing changes between states. 
The vocabulary for a transition is given by a \emph{transition signature} $\Theta$ of the form 
$(\tuple{\Dyn},\\ \tuple{ \pt{\Dyn}}, \tuple{\Stat})$\@. $\tuple{\Dyn}$ and $\tuple{\pt{\Dyn}}$ (for \emph{Dynamic}) are tuples of relational symbols. $\tuple{Dyn}$ and $\pt{\tuple{Dyn}}$ are similar: they have matching length and pointwise arities. $\tuple{\Stat}$ (for \emph{Static}) is a tuple of symbols. 

$\tuple{\Dyn}$ provides the vocabulary for describing the precondition, $\tuple{\pt{\Dyn}}$ for describing the postcondition, $\tuple{\Stat}$ for describing the invariant part,
and $\form P$, $\pt{\form P}$ for describing the elements \emph{present} in the pre- and postconditions. 

$\tuple{\Dyn}$ contains a distinguished unary predicate symbol $\form{P},\pt{\form P} \in \tuple{\Dyn}$ (for \emph{Presence}) so node creation/destruction can be encoded.  
The coherence constraint $\textit{Support} \dfn \fa{x} \form{P}(x) \vee \pt{\form{P}}(x)$
prevents spurious nodes that would inhabit a structure yet be encoded as nonexistent. $\mdl A$ is \emph{supported} if $\mdl A \vDash \textit{Support}$ and $\phi$ is a \emph{support formula} if $\phi\vDash\textit{Support}$.

A formula $\phi$ is \emph{pre} if it does not use any symbol from $\pt{\tuple{\Dyn}}$\@. If $\phi$ is a formula, $\pt \phi$ is $\phi$ where every symbol from $\tuple{\Dyn}$ has been replaced by its counterpart from $\pt{\tuple{\Dyn}}$\@.


\section{Transitions, Forests of Linked Bounded trees}\label{s:transitions_flbs}

An \emph{FLB} signature (for \emph{Forest of Linked Bounded trees}) is a transition signature specialised for representing transitions on bounded forests with dynamic links between nodes. There is a convenience \emph{parent} function symbol which goes up one level in the tree. Other function names play the role of tree edge labels.  Nodes can be linked through a functional and symmetric relation $\form{Link}$ (or $\pt{\form{Link}}$ in the postcondition).\footnote{One can increase the number of binding partners by allowing a subtree under a node.}

A transition signature $\Sigma \dfn (\tuple{\Dyn},\tuple{\pt{\Dyn}},\tuple{\Stat})$ is an \emph{FLB signature} whenever :
\begin{align*}
  \tuple{{\Dyn}} &= \form{P}, \form{Link}, \tuple{L} &
  \tuple{\pt{\Dyn}} &= \pt{\form{P}},\pt{\form{Link}}, \pt{\tuple{L}} &
  \tuple{\Stat} &= \tuple{f}, \textit{parent}, \tuple{N} \span\span
\end{align*}
where $\form{P},\pt{\form P}$ are the unary presence symbols. $\form{Link},\pt{\form{Link}}$ are binary link symbols. $\tuple{L}$ (dynamic \emph{Labels}) is a tuple of unary predicate symbols. $\tuple f$ is a tuple of unary functions (child-of functions). $|\tuple f|$ bounds the degree of the trees. $\textit{parent}$ is a distinguished unary function. $\tuple{N}$ (static \emph{names}) is a tuple of unary predicate symbols.

Not all $\Sigma$-structures are forests of linked bounded trees. 
With $\mdl A$ a $\Sigma$-structure, $G_{\mdl A}$ is the loop-free\footnote{For any $x$, $(x,x)$ is not in the graph even if $f(x)=x$\@.} union of graphs of
$\{\sem{f}{\mdl A}\ |\ f \in \tuple f\}$\@.
$\mdl A$ is an FLB whenever $G_{\mdl A}$ is a forest, the loop-free graph of $\sem{\textit{parent}}{\mdl A}$ is the parent relation in that forest, 
and $\sem{\form{Link}}{\mdl A},\sem{\pt{\form{Link}}}{\mdl A}$ are symmetric and functional.

For every $n \geq 0$, the $n$-FLBs are the FLBs with trees of height at most $n$\@. 

\begin{example}
In figure \ref{f:iota_example_bad}, $\mdl T'$ is a $2$-FLB. The symbols Raf-1 and Tyr are in $\tuple N$ of the underlying signature. $\form{Phos}$ is in $\tuple L$ and $\pt{\form{Phos}}$ is in $\pt{\tuple L}$\@. 
There is a function symbol $f \in \tuple f$ such that $\sem{f(parent(m_1))}{\mdl T'} = m_1$\@.
The creation of a link between $m_1$ and $m_2$ is encoded as $(m_1,m_2) \notin \sem{\form{Link}}{\mdl T'}$ and $(m_1,m_2) \in \sem{\pt{\form{Link}}}{\mdl T'}$\@.
The creation of $m_3$ is encoded as $m_3 \notin \sem{\form P}{\mdl T'}$ and $m_3 \in \sem{\pt{\form P}}{\mdl T'}$\@.
\end{example}

$n$-FLBs can be characterised by a finite, first-order, universal\footnote{That is, $\Tn$ is in the $\forall^*$ prenex class.} theory $\Tn$\@. We do not reproduce it here in full detail\extendedversion{ (see Appendix)}. It is of the following form:
\[
  \Tn \dfn 
  \fa{x} \iform{ParentSpec}(x) \wedge 
  H^n(x) \wedge
  \textit{FunSymLink} \wedge 
  \pt{\textit{FunSymLink}}
\]
$\iform{ParentSpec}$ forces $\textit{parent}$ to behave as a parent function. $H^n$ forces paths through $\tuple f$ to be of length at most $n$.\footnote{Note that the signature bounds the degree of the trees, while the theory bounds their height.} $\textit{FunSymLink}$ and $\pt{\textit{FunSymLink}}$ ensure that $\form{Link}$ and $\pt{\form{Link}}$ are functional and symmetric. 
\begin{lemma}\label{l:flb+tn}
  A $\Sigma$-structure $\mdl A$ is an $n$-FLB iff $\mdl A \models \Tn$\@.
\end{lemma}

The proof uses $H^n$ to prevent cycles and $\iform{ParentSpec}$ to force unicity of paths from the roots.

Formulas modulo $\Tns$ are a good candidate for knowledge representation, but querying is not possible in general. Let $\Tns \dfn \Tn \wedge \textit{Support}$ be the theory of supported $n$-FLBs: 

\begin{theorem}\label{l:undecidable}
  First-order satisfiability modulo $\Tns$ is undecidable for $n \geq 1$\@.
\end{theorem}
The proof is by reduction of domino problems \extendedversion{(see Appendix)}. Colors are labels in $\tuple N$, and trees have height $1$, colored roots, and $4$ leaves. Each leaf is a direction (up,down,left,right) and the only allowed links are between up-down or left-right pairs with appropriately colored roots.

For any FLB signature, satisfiability modulo $\Tns$ is still achievable in a restricted fragment:
\begin{theorem}\label{p:l:tns+decidable}
  For $n \geq 0$, satisfiability modulo $\Tns$ in the $\exists^*\forall^*$ fragment is decidable.
\end{theorem}
This can be proved by adapting the classic proof of decidability for the Bernays-Sch{\"o}nfinkel-Ramsey fragment in relational FO with equality to our non-relational signatures. We show that $\Tn$ restrains functions enough to maintain decidability because iterated function application becomes stationary after a bounded number of steps. As in the original proof, we get a small model property as a byproduct and a description of that model (it has just enough trees to host the existential witnesses required by the $\exists^*$ part).


FLB signatures and their associated theories $\Tns$ describe state transitions on forests of bounded trees with static and dynamic labels as well as a dynamic, functional link relation between nodes. While satisfiability is not decidable in general, it is in the $\exists^*\forall^*$ fragment. Note in our example that $\textit{Obs}(x) \in \exists^*\forall^*$\@. The next section introduces commonsense reasoning by characterising transitions which, given a precondition, only apply the changes that are necessary to satisfy a formula.
\section{Change minimisation}\label{s:change_minimisation}
%
%
%
For $\form A \in \tuple{\Dyn}$, 
and $\tuple x$ an arity($\form A$)-sized tuple of variables,
$\Delta\form A(\tuple x)$ describes the changes in $\form A$:
$\Delta \form A(\tuple x) \dfn \form{A}(\tuple x) \leftrightarrow \neg \pt{\form{A}}(\tuple x)$\@.
For simplicity, the tuple $\tuple x$ may be omitted.

$\Theta$-structures can be ordered along a partial \emph{change order} $\actordeq$: for $\mdl A, \mdl B$ any two $\Theta$-structures, let $\mdl A \actordeq \mdl B$ whenever for all $\form A \in \tuple{\Dyn}$:
%
\begin{align*}
  \sem{\tuple{\Dyn}}{\mdl A} &= \sem{\tuple{\Dyn}}{\mdl B} &
  \sem{\tuple{\Stat}}{\mdl A} &= \sem{\tuple{\Stat}}{\mdl B}\restriction \dom(\mdl A)\\
  \sem{\Delta\form A}{\mdl A} &\subseteq \sem{\Delta\form A}{\mdl B} &
  \dom(\mdl A) &\subseteq \dom(\mdl B)
\end{align*}
So $\mdl A \actordeq \mdl B$ means that they have equal preconditions, that $\mdl B$ contains at least the elements in $\mdl A$, and that any change that occurs in $\mdl A$ also occurs in $\mdl B$.

%
%

\begin{example}
Consider $\mdl T$ and $\mdl T'$ from figures \ref{f:simple_illustration} and \ref{f:iota_example_bad}. $\mdl T \actord \mdl T'$: their precondition ($\tuple{\Dyn}$) are equal, their static parts ($\tuple{\Stat}$) are equal on their common elements, 
$\mdl T'$ has one more element ($m_3$) and, 
while every change in $\mdl T$ is present in $\mdl T'$, 
$m_1 \in \sem{\Delta\form{Phos}}{\mdl T'}$ but $m_1 \notin \sem{\Delta\form{Phos}}{\mdl T}$\@. %
\end{example}

The $\actordeq$-minimal models of a first-order formula $\phi$ are expressed as $\circum\phi$ (``minimised $\phi$''):

\begin{definition}
  With $\phi$ a formula, $\mdl A,\mu \models \circum\phi$ iff $\mdl A,\mu\models\phi$, and there is no $\mdl B \actord \mdl A$ such that $\mdl B,\mu\models\circum\phi$.
 
\end{definition}


\begin{example}
  Compare $\mdl T$ in figure \ref{f:simple_illustration} and $\mdl T'$ in figure \ref{f:iota_example_bad}. Both satisfy $\textit{Obs}(m_1)$\@. But $\mdl T \actord \mdl T'$, so $\mdl T'$ does not satisfy $\circum\textit{Obs}(m_1)$\@.
\end{example}

Intuitively, if $\phi$ represents existing knowledge of a biological mechanism, $\circum\phi$ represents the current best model (in the biological sense) implied by that knowledge. 

One may naturally ask for a syntactic definition of $\downarrow$. In section \ref{s:circumscription}, we will see that, in general, $\circum\phi$ is second-order expressible \extendedversion{(see Appendix for the full proof)}.  
In the meantime, the next section provides deduction rules that can produce formulas of the form $\circum\phi$. It defines a class of formulas with minimal models that can be captured in a first-order fragment, rather than in second-order logic only.



\section{Deduction rules}\label{s:deduction_rules}

We introduce deduction rules for the judgement $\vdash$, which should be seen as a typing property for formulas. 

For any term $u$, $\alpha(t,u)$ is any binary atom where $t$ and $u$ both appear.
If $\tuple T$ is a tuple of relational symbols, $\mc L_{\tuple T}$ is the set of literals that use symbols of $\tuple T$\@.

$\safe$ is any set of first-order variables, $d \geq 0$, and $\phi$ is a formula. In a judgment of the form $\safe\consep d \vdash \phi$, we say that $\safe\consep d$ is the context. Functions of FLB signatures are unary, so for any term $t$, $\varof t$ is a singleton $\set{x_t}$ and for $\pi \in \set{\forall,\exists}$, $\pi \varof t \dfn \pi x_t$\@.

The judgment $\safe\consep d \vdash \phi$ implies that, for any $\mdl A \models \Tns$ and $\mu : \varof\phi\rightarrow\dom(\mdl A)$, there is a ``protected'' subset $S \subseteq \dom(\mdl A)$ parameterized by $\safe$, $d$ and $\mu$ such that removing changes of $\mdl A$ outside of $S$ preserves satisfaction of $\phi$ (see section \ref{s:preservation}).
 \newcommand{\mquad}{}
 \newcommand{\ruleWeak}{%
 \begin{prooftree}
    \AxiomC{$\safe \consep d \vdash \phi $}
    \LeftLabel{\parbox[c]{2.5em}{
        {\scriptsize
          $\safe \subseteq \safe'$\\[-0.4em]
        $d \leq d'$
      }
    }}
    \RightLabel{\mquad\intro{rule:weak}}
    \UnaryInfC{$\safe' \consep d'\vdash \phi$}
  \end{prooftree}%
}

\newcommand{\ruleBool}{%
  \begin{prooftree}
    \AxiomC{$\safe\consep d \vdash \phi_1$}
    \AxiomC{$\safe\consep  d \vdash \phi_2$}
    \LeftLabel{\parbox[c]{4em}{
        {\scriptsize
          $\oplus \in \set{\wedge,\vee}$
        }
    }}
    \RightLabel{\mquad\intro{rule:bool}}
    \BinaryInfC{$\safe \consep d\vdash \phi \oplus \phi_1$}
  \end{prooftree}%
}

\newcommand{\ruleForall}{%
  \begin{prooftree}
    \AxiomC{$\safe\consep d \vdash \phi$}
    \AxiomC{$\safe'\consep \_ \vdash \alpha(t,u)$}
    \LeftLabel{
      \scriptsize%
      {$\varof t \neq \varof u \cap (\safe\cup\safe')$}
    }
    \RightLabel{\mquad\intro{rule:forall}}
    \BinaryInfC{$\addrem{\safe}{\varof u}{\varof t}\consep d + |\varof t \cap \safe| \vdash \fa{\varof{t}} \alpha(t,u) \rightarrow  \phi$}
  \end{prooftree}%
}

\newcommand{\ruleExists}{%
  \begin{prooftree}
    \AxiomC{$\safe\consep d \vdash \phi$}
    \AxiomC{$\safe'\consep \_ \vdash \alpha(t,u)$}
    \LeftLabel{
    \scriptsize
      \mbox{
        {$\varof{t} \neq \varof{u}$}
    }}
    \RightLabel{\mquad\intro{rule:exists}}
    \BinaryInfC{$\addrem{\safe}{\varof u}{\varof t} \consep d+1\vdash \ex{\varof{t}} \alpha(t,u) \wedge  \phi$}
  \end{prooftree}%
}

\begin{note}
  The distance is always increased by $1$ because we want to be semantically correct. otherwise, we could have $d + |\varof t \cap (\safe \cup \safe')|$. Why? First consider $\pt R(x,y) : (y,1)$. Without the system $+1$ increase, we would wrongly apply the rule and get $\pt R(x,y) : (x,0)$. If we only look at the syntax, we can't prove that this atom is $(y,1)$, just that it's the case for the equivalent $\ex{z}\pt R(x,z) \wedge x = y$. but semantically the rule would be wrong.

  Also we dont just do $d + |\varof t \cap \safe|$ because $\pt R(x,y) : (xy,0)$, and this needs a distance-1 protection (unlike $\neg \pt R(x,y)$, because of the special definition of sub).
\end{note}

\newcommand{\ruleInv}{%
  \begin{prooftree}
    \AxiomC{$\set{x}\consep 0 \vdash \pt\phi$}
    \LeftLabel{
      {\scriptsize
        $\phi$ pre
    }}
    \RightLabel{\mquad\intro{rule:inv}}
    \UnaryInfC{$\emptyset\consep 0\vdash\phi \wedge \pt \phi$}
  \end{prooftree}%
}

\newcommand{\ruleDynamic}{%
  {\begin{prooftree}
    \AxiomC{}
    \LeftLabel{
      {\scriptsize
        $L \in \mc{L}_{\pt{\tuple P}, \pt R}$
    }}
    \RightLabel{\mquad\intro{rule:dynamic}}
    \UnaryInfC{$\varof{L}\consep 0 \vdash L$}
\end{prooftree}}
}

\newcommand{\ruleStatic}{%
  \begin{prooftree}
    \AxiomC{}
    \LeftLabel{
      {\scriptsize
        $L \in \mc{L}_{\tuple P,\tuple N, R, =}$
    }}
    \RightLabel{\mquad\intro{rule:static}}
    \UnaryInfC{$\emptyset\consep  0 \vdash L$}
  \end{prooftree}%
}

%
\newcommand{\ruleCircum}{%
\begin{prooftree}
  \AxiomC{$\safe\consep d \vdash \phi$}
  \RightLabel{\mquad\intro{rule:circum}}
  \UnaryInfC{$\varof{\phi}\consep d \vdash \circum(\phi \wedge \Tns)$}
\end{prooftree}%
}

\noindent%
\begin{minipage}[c][4em]{0.5\textwidth}%
  \ruleDynamic
\end{minipage}
\begin{minipage}[c][4em]{0.5\textwidth}
  \ruleStatic
\end{minipage}
\begin{minipage}[c][4em]{0.4\textwidth}%
  \ruleWeak
\end{minipage}
\begin{minipage}[c][4em]{0.6\textwidth}
  \ruleBool
\end{minipage}
\begin{minipage}[c][4em]{0.5\textwidth}%
  \ruleCircum
\end{minipage}
\begin{minipage}[c][4em]{0.5\textwidth}
  \ruleInv
\end{minipage}
\begin{minipage}[c][4em]{1\textwidth}
  \ruleForall
\end{minipage}
\begin{minipage}[c][4em]{1\textwidth}
  \ruleExists
\end{minipage}

\begin{note}
    For the GUARD rules, there are many parameters to play with:
    \begin{itemize}
      \item It is imperative that the rules are semantically correct; that is, that the conclusion follows from the premises when $\safe\consep d \vdash \phi$ is interpreted as ``$\phi$ is preserved under $\safe\consep d$''. This applies to $\alpha(t,u)$ in particular: I don't want a rule that works only because $\set{x}\consep d \vdash \pt R(x,y)$ is not derivable even though it is semantically the case that we can protect $x$ with a radius of $1$.
      \item However I can choose to define $\alpha(t,u)$ syntactically OR semantically; that is, to say: ``$\alpha(t,u)$ \emph{is} an atom'', or to say ``$\alpha(t,u)$ is \emph{equivalent} to an atom''. In particular, defining $\alpha$ modulo equivalence lets me use $\set{x}\consep 1 \vdash \pt R(x,y)$ because an equivalent formula with this context can be derived using "rule:exists" (while syntactically it is not derivable). In a nutshell, I have to \emph{allow} formulas equivalent to an $\alpha$, but I can choose to \emph{force} the proof system to accept them, or not. Forcing it lets me write a simpler rule: I don't have to have separate $\safe$ and $\safe'$ in "rule:forall" anymore. So basically the choice is whether I go into \emph{deduction modulo}, or not.
      \item The other choices are whether I have $+1$ or $+|\varof t \in \safe|$ in the conclusion of "rule:exists"; whether $\safe = \safe'$ or not; whether $d = d'$ or not. All those choices, and the ones before, are interconnected in subtle ways.
    \end{itemize}

 It's not enough that the rules are correct! They're also not allowed to \emph{lose} derivations they had before (up to equivalence)! 

    As an example, here is a pair of rules that work (I think):

  \begin{prooftree}
    \AxiomC{$\safe\consep d \vdash \phi$}
    \AxiomC{$\safe\consep \_ \vdash \psi$}
    \LeftLabel{
      \scriptsize%
      \Centerstack[c]{{$\varof t \neq \varof u \cap \safe$}
      $\psi \equiv \alpha(t,u)$}%
    }
  \RightLabel{Aguard'}
    \BinaryInfC{$\addrem{\safe}{\varof u}{\varof t}\consep d + |\varof t \cap \safe| \vdash \fa{\varof{t}} \alpha(t,u) \rightarrow  \phi$}
  \end{prooftree}

  \begin{prooftree}
    \AxiomC{$\safe\consep d \vdash \phi$}
  \AxiomC{$\safe\consep d \vdash \psi$}
    \LeftLabel{
    \scriptsize%
    \Centerstack[c]{
        {$\varof{t} \neq \varof{u}$}
        $\psi \equiv \alpha(t,u)$
  }}
    \RightLabel{Eguard'}
    \BinaryInfC{$\addrem{\safe}{\varof u}{\varof t} \consep d+|\varof t \cap \safe|\vdash \ex{\varof{t}} \alpha(t,u) \wedge  \phi$}
  \end{prooftree}

  In the last rule, I could also have $d+1$ and $\safe\consep -$ instead of $\safe\consep d$. Either way, this gives us a more elegant "rule:forall", and a pair of guarded rules that are more similar: the only differences are the proviso on $\varof t$, and either one difference in the premise or one difference in the conclusion.
\end{note}

\begin{note}
    \begin{itemize}
      \item By forcing $+1$ in "rule:exists", I remove $\fa{x}\ex{y} R(x,y) : (\emptyset,0)$. This could theoretically exist in invariants etc, but AFAICT it takes us out of $\exists^*\forall^*$. I may as well force $+1$ and have every derivable formula... edit: wait do I need that? I already add x to $\safe$ when I say $\ex{y} R(x,y)$, so I don't think I can add a lone $\forall$ in front of that. In which case maybe I can add $[\varof t \in \safe]$, with $\safe = \safe'$; and I can also go to $\safe = \safe'$ for "rule:forall"... this adds symmetry to the whole thing.
      \item With a generalized "rule:bool" ($\safe_1 \cup \safe_2\consep \max(d_1,d_2)$), weakening for $\safe$ is derivable ($\pt R(x,x) \vee \neg \pt R(x,x)$), and to derive weakening for $d$, I probably have to use some version of "rule:exists".
      \item I will stick with a multiplicative version because 
        1) the additive version doesn't fully unify "rule:exists" and "rule:forall" for reasons 
        a) I want to force $+1$ on "rule:exists", see above, and 
        b) I can't force the same $d$ on both sides in "rule:forall" 
        otherwise I can't derive $\pt R(x,y) : (y,0)$, also 2) later I want to say as a strategy ``don't use "rule:exists" in invariants''. To do that I need $\pt R(x,y) : (y,0)$ derivable without use of "rule:exists", otherwise I'd have to say ``first add a special gadget, you're allowed to use this gadget...'' boring.
    \end{itemize}
\end{note}


We state the main theorem of the paper and informally describe the rules. The remaining sections introduce the main theoretical tools that are necessary to prove the theorem.
\begin{theorem}\label{l:main+theorem}
  If $\safe\consep d \vdash \phi$, then $\phi \wedge \Tns \in \exists^*\forall^*$ and $\phi \in \forall^*\exists^*$\@.
\end{theorem}

Proposition \ref{p:l:tns+decidable} and Theorem \ref{l:main+theorem} imply that, modulo $\Tns$, validity and satisfiability are decidable for $\vdash$-deducible formulas. In particular, consider the rule "rule:circum", which has no special proviso. Any deducible formula can be minimised along $\actordeq$ (modulo $\Tns$), and the result is not only first-order expressible, but also equivalent both to a formula in $\exists^*\forall^*$ and to one in $\forall^*\exists^*$\@.

"rule:static" and "rule:dynamic" both introduce literals, but "rule:dynamic", being about the postcondition (note the proviso $L \in \mathcal{L}_{\tuple{\pt p},\pt R}$), must protect the elements mentioned in $L$\@. "rule:weak" says that the protected area can always be expanded. "rule:bool" says that boolean combinations are allowed. "rule:inv" says that, if the protected area is small enough, it can be ignored as long as constraints on the postcondition are extended to the precondition. While "rule:bool" and "rule:inv" may both produce new conjunctions, "rule:inv" can remove an element from the protected set provided additional constraints are satisfied. "rule:forall" and "rule:exists" introduce quantifiers. The proviso for "rule:exists" requires a proper guard $\alpha(t,u)$ ($\varof t \neq \varof u$) and increases the protection distance $d$ by $1$. The proviso for "rule:forall" allows a vacuous guard ($\varof t = \varof u$) in some cases, and does not always increase the protection distance. The asymmetry between "rule:forall" and "rule:exists" reflects the asymmetry in the notion of ``protection'', cf. section \ref{s:preservation}.

\begin{example}
  $\textit{Obs}(x)$ and $\circum(\textit{Obs}(x) \wedge \Tns)$ are deducible. For instance, $\set{x}\consep 0 \vdash \textit{Observation}_1(x)$ is derived by applying "rule:forall" to $\neg\pt{\form{Active}}(\textit{parent}(x))$ as $\phi$  and $\pt{\form{Link}}(x,y)$ as $\alpha$ (both introduced with "rule:dynamic").
\end{example}

\section{Proof elements for Theorem \ref{l:main+theorem}}\label{s:proof_sketch}

We focus on techniques with general applicability. 
%
%
Subsection \ref{s:preservation} introduces \emph{preservation}, the main semantic invariant which is implied by $\vdash$\@. Preservation captures a notion of constraint locality at the semantic level which then translates to syntactic expressivity properties.
Subsections \ref{s:circumscription} and \ref{s:circum_application} detail how the operator $\circum$ is constructed as an instanciation of \emph{unified circumscription}, a generalisation of existing circumscription schemes. Subsection \ref{s:main+theorem} sketches how preservation implies first-order expressibility of circumscribed formulas modulo $\Tns$ and why the resulting first-order formula lives in both $\exists^*\forall^*$ and $\forall^*\exists^*$\@.
\subsection{Preservation}\label{s:preservation}

The intuition behind preservation is to find classes of formula that provide useful static information on their $\actordeq$-minimal models. In particular, it implies that changes in minimal models are in a ball of bounded radius, which lets them be characterised by first-order formulas. Preservation also interacts well with formula composition.

Let $\Sigma$ be an FLB signature. For $\form A \in \tuple{\Dyn}$, let 
  $\added \form A(\tuple x) \dfn \Delta\form A(\tuple x) \wedge \neg \form A(\tuple x)$, and
  $\removed \form A(\tuple x) \dfn \Delta\form A(\tuple x) \wedge \form A(\tuple x)$\@.
Let $\mdl A$ be an FLB for $\Sigma$.
$T_{G_{\mdl A}}$ is the set of trees of $G_{\mdl A}$\@. For $t \in T_{G_{\mdl A}}$, $V_t$ is the set of vertices of $t$\@. For $a \in \dom(\mdl A)$, $t_a \in T_{G_{\mdl A}}$ is the tree such that $a \in V_{t_a}$\@.


\begin{definition}
  A node $a \in \dom(\mdl A)$ is \emph{modified} whenever at least one of the following is true:\\
  -- $a \in \sem{\Delta \form A}{\mdl A}$ for some unary $\form A \in \tuple{\Dyn}$\\
  -- There is $b \in \dom(\mdl A)$ such that $(a,b) \in \sem{\added\form{Link}}{\mdl A}$\\
  -- There is $b \in V_{t_a}$ such that $(a,b) \in \sem{\removed \form{Link}}{\mdl A}$
\end{definition}

In particular, an external link deletion (some $(a,b) \in \sem{\removed \form{Link}}{\mdl A}$ with $b \notin V_{t_a}$) does not make $a$ a modified element.
A tree $t$ is \emph{modified} whenever at least one of its elements is modified. The set of modified elements in $\mdl A$ is denoted by $\mc C(\mdl A)$.
For any tree $t$, the set of \emph{modified elements outside $t$} is 
$\changed{t}(\mdl A) \dfn \mc C(\mdl A)\setminus V_t$.


\begin{definition}
For any nodes $a,b \in \dom(\mdl A)$ the \emph{link distance} $d_{\mdl A}(a,b)$ is the distance between $t_a$ and $t_b$ in the graph with nodes $T_{G_{\mdl A}}$ and edges $\set{(t_c,t_d)\ |\ (c,d) \in \sem{\form{Link}}{\mdl A}\cup\sem{\pt{\form{Link}}}{\mdl A}}$\@. 

For $d\geq 0, K \subseteq \dom(\mdl A)$, the \emph{ball of radius $d$ around $K$} is:
\[
  \ball_{\mdl A}(K,d) \dfn \set{a\ |\ \min_{b\in K} d_{\mdl A}(a,b) \leq d}
\]
\end{definition}

If we protect a ball of radius $d$ around a set $K \subseteq \dom(\mdl A)$, we can \emph{clear} the changes of a tree $t$ outside of that protected area and produce a new FLB $\mdl B$. Intuitively, we:

  \begin{enumerate}
    \item Pick a tree $t$ far enough (at distance $d$) from a special set ($K$), then 
    \item Clear any modification that relates to $t$, and
    \item Clear external edge deletions that relate to $t$ and unprotected, unmodified trees.
  \end{enumerate}

  \begin{definition}
Let $R$ be a relation and $X$ a set, $R \restriction X$ are the tuples of $R$ that mention at least one element of $X$\@. $R-X$ are the tuples of $R$ that mention no element of $X$\@.
\end{definition}
\begin{definition}
With $K \subseteq \dom(\mdl A)$, $d\geq 0$,
$t$ a tree of $G_{\mdl A}$ that does not intersect $\ball_{\mdl A}(K,d)$, we say that $\mdl B \actordeq \mdl A$ is a \emph{$(K,d)$-sub of $\mdl A$ with cleared tree $t$} whenever, for all $\form{A}\in\tuple{\Dyn}$: 
  \begin{flalign*}
    \sem{\Delta \form A}{\mdl B} &= \sem{\Delta \form A}{\mdl A} \setminus V_t \hspace{4em}
    \sem{\added \form{Link}}{\mdl B} = \sem{\added \form{Link}}{\mdl A} - V_t\\
    \sem{\removed \form{Link}}{\mdl B} &= (\sem{\removed \form{Link}}{\mdl A} - V_t) \cup (\sem{\removed \form{Link}}{\mdl A} \restriction (\changed{t}(\mdl A) \cup \ball_{\mdl A}(K,d)))
  \end{flalign*}
\end{definition}

  If $t$ is not specified, we say that $\mdl B$ is a $(K,d)$-sub of $\mdl A$\@. If $(K,d)$ is not specified, we say that $\mdl B$ is a sub of $\mdl A$\@. Note that the resulting sub is not uniquely defined (elements of the domain may disappear). 

\begin{example}
We illustrate subs in figure \ref{f:sub_example}. We assume no changes in $\tuple L$\@. The pre- and postconditions are superimposed: there are changing links between $t_1, t_2, t_3, t_4$ with a solid link for an addition, and a dashed one for a deletion (no link is both in the pre- and postcondition). The effect of going from $\mdl A$ to a $\{a\}$-sub $\mdl B$ of $\mdl A$ with cleared tree $t_1$ is illustrated by the red areas that indicate which link changes are cleared (i.e. are in $\sem{\Delta \pt{\form{Link}}(x,y)}{\mdl A}$ but not in $\sem{\Delta \pt{\form{Link}}(x,y)}{\mdl B}$). The striped area is $\ball_{\mdl A}(\{a\},0)$\@. The link deletion from $t_1$ to $t_4$ is not cleared, because it touches a node in the tree of the kernel $\{a\}$; neither is the link deletion between $t_1$ and $t_3$ because $t_3$ is \emph{changing} (both through a link addition with $t_1$ and an internal link deletion). However, the link addition between $t_1$ and $t_3$ is cleared (unconditionally), as well as the internal link deletion on $t_1$ since, even though $t_1$ is changing, it is also the cleared tree and thus unprotected. Finally, the link between $t_1$ and $t_2$ is cleared because $t_2$ is neither changing nor in $\mc B_{\mdl A}(\{a\})$\@.
\begin{figure}
  \centering
  \caption{$\{a\}$-sub with cleared tree $t_1$}
  \label{f:sub_example}
  \includegraphics[width=0.80\textwidth]{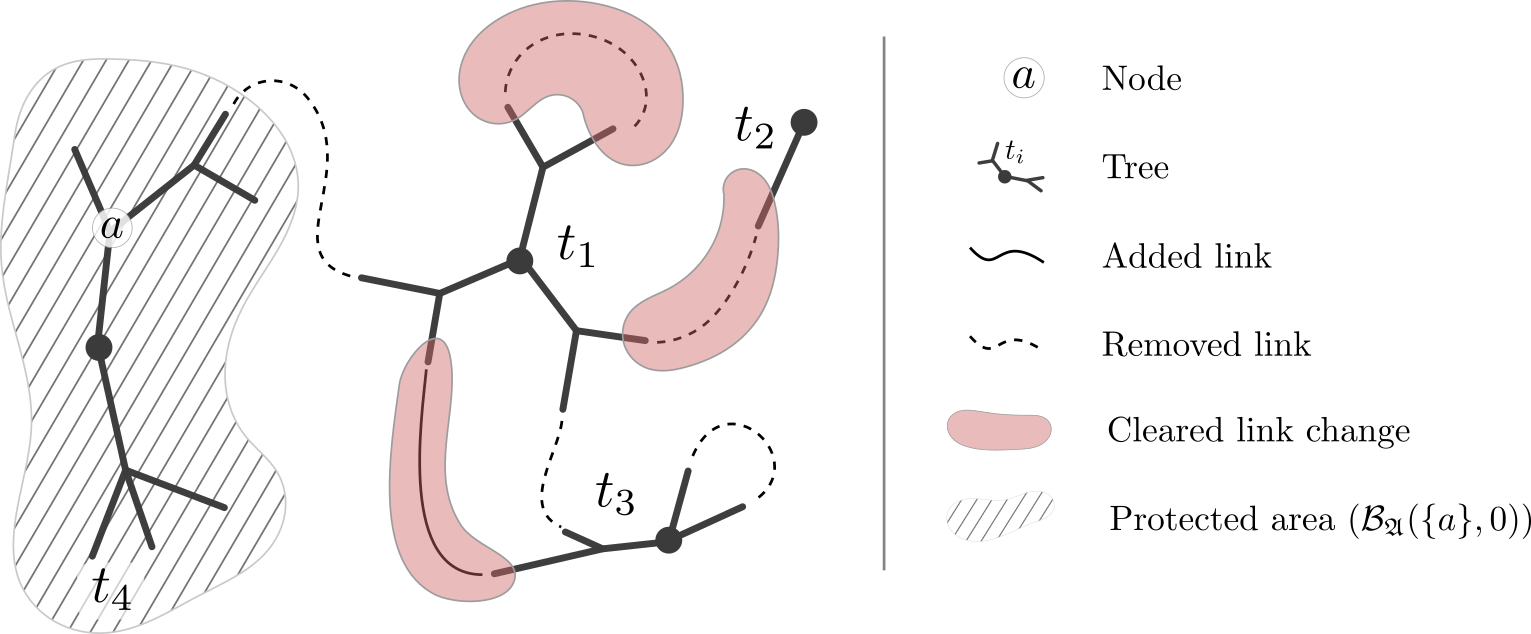} 
\end{figure}
\end{example}

The idea is that, for a class of formulas, satisfaction is preserved by taking subs.
If a change (unary predicate change, or edge deletion, or edge addition) is present in $\mdl A$ but not present in $\mdl B$, we say that it has been \emph{cleared}.
This new relation between structures induces a property on formulas we call \emph{preservation}:


\begin{definition}\label{def:preserved}
For $d\geq 0$, $\safe$ a set of variables, a formula $\phi$ is \emph{preserved under $\safe\consep d$} if for all FLBs $\mdl A$, all $\mu : \varof{\safe} \rightarrow \dom(\mdl A)$, all $(\sem{\safe}{\mdl A,\mu},d)$-subs $\mdl B$ of $\mdl A$, all $\nu : \varof{\phi}\!\setminus\! \varof{\safe} \rightarrow \dom(\mdl B)$, $\mdl A, \mu, \nu \vDash \phi$ implies  $\mdl B, \mu, \nu \vDash \phi$\@.
\end{definition}

\begin{theorem}\label{l:proved+preserved}
  If $\safe\consep d\vdash\phi$ then $\phi$ is preserved under $\safe\consep d$\@.
\end{theorem}

We give a proof sketch for each rule:

\textbf{"rule:dynamic"}: Take $\pt{\form A}(a)$ as an example. In an FLB $\mdl A$, the only clearing of changes that could invalidate $a \in \sem{\pt{\form A}}{\mdl A}$ would be the clearing of $t_a$\@. If $\mdl A \vDash \pt{\form A}(a)$, $t_{a} \in \ballA(\set{a},0)$, and so $t_{a}$ can never be the cleared tree.

\textbf{"rule:static"}: Constraints on the precondition, on equality or on static properties can not be invalidated by clearing changes, as that only modifies postconditions. Taking subs protects elements in the image of the interpretation of variables.

\textbf{"rule:weak"}: Taking a larger protected area (either by adding elements to $\safe$ or by increasing $d$) can only protect more trees.

\textbf{"rule:bool"}: We explain for $\oplus = \vee$\@. Consider a $(\sem{\safe}{\mdl A,\mu})$-sub $\mdl B$ of $\mdl A$: if $\mdl A,\mu,\nu \vDash \phi_i$, then by hypothesis, so does $\mdl B,\mu,\nu$\@.

\textbf{"rule:circum"}: Minimising a formula \emph{modulo $\Tns$} leaves only models that have no strict $(\sem{\safe}{\mdl A,\mu},d)$-subs, so the claim becomes vacuously true. 

\textbf{"rule:inv"}: Consider for example $\pt{\form A}(a)$: as shown with "rule:dynamic", $a$ must be protected. More precisely, suppose $A(a)$ is false in $\mdl A$ and $\pt{\form A}(a)$ is true. $\pt{\form A}(a)$ can be made false by clearing the wrong tree. Consider $\form A(a) \wedge \pt{\form A}(a)$\@. Clearing changes on $t_{a}$ may no longer invalidate the formula. The rule "rule:inv" extends this reasoning to first-order specifications that require a single protected element. 

\textbf{"rule:forall"} and \textbf{"rule:exists"}: The important aspect of quantification is that a variable becomes ``hidden'' from $\safe$\@. If the interpretation of a variable had to be protected, the new context must ensure that the protection remains, even once the variable has become unreachable. $\alpha(t,u)$ functions as a guard: it links $t$ to $u$ and adds $\varof u$ to the context. 

There are two differences between "rule:forall" and "rule:exists" which make "rule:forall" more relaxed. 

First, the proviso $\varof t \neq \varof u$ in "rule:exists" excludes formulas such as $\ex{x} x=x \wedge \text{N}(x)$ (with $N \in \tuple{\Stat}$). Taking the sub of an FLB may remove elements from the domain, so the existence of an element satisfying a static property is never guaranteed under subs. In "rule:forall", $\varof t = \varof u$ is possible as long as $\varof t$ needs no protection (i.e. $\varof t \notin \safe \cup \safe'$), because a universal quantifier is not invalidated by domain reduction.

Second, the protection distance is systematically increased by $1$ in the case of "rule:exists", but not in the case of "rule:forall". For instance  
$\pt{\form{Link}}(x,y)$ is preserved under $\set{x,y}\consep 0$, but $\ex{y} \pt{\form{Link}}(x,y)$ is preserved under $\set{x}\consep 1$, not under $\set{x}\consep 0$: if a link between (the images of) $x$ and $y$ is created, clearing changes in $y$'s tree wil unconditionally clear that link creation, thus invalidating the formula. So we either need to protect $y$ directly, or we need to protect a ball of radius at least $1$ around $x$\@. In the case of "rule:forall", the asymmetry in the definition \emph{modification} is exploited: link deletions to protected trees may not be cleared, so it is not necessary to extend the protection radius. For instance, $\neg \pt{\form{Link}}(x,y)$ is preserved under $\set{x}\consep 0$\@.

Preservation becomes useful when one considers $\actordeq$-minimal models of a preserved formula. First, we need to make the definition of $\downarrow$ explicit.

\subsection{Unified Circumscription}\label{s:circumscription}

Circumscription is an umbrella term for second-order characterisations of the minimal models of a first-order formula $\phi$ along an order. We combine general domain circumscription (GDC) \cite{Doherty1998,McCarthy1980} and parallel predicate circumscription \cite{McCarthy1986}. 

Any signature $\Upsilon$ is partitioned into tuples of predicates and functions: 
\[
  \Upsilon \dfn (\tuple P_\fx, \tuple P_\vr, \tuple P_\rs, \tuple f_\rs, \tuple P_\mn)
\]
As in both GDC and parallel circumscription, some predicates are varying ($\tuple P_\vr$). As in GDC, the domain is circumscribed and some predicates and functions are fixed on the restricted domain ($\tuple P_\rs, \tuple f_\rs$). As in parallel circumscription, some predicates are circumscribed ($\tuple P_\mn$) and others are fixed on the initial domain ($\tuple P_\fx$).\footnote{For simplicity, we omit varying and fixed functions from the definition (not necessary here).}

Such a partition on $\Upsilon$ induces an associated order: for $\mdl A,\mdl B:\Upsilon$, $\mdl B \circordeq \mdl A$ whenever
\begin{align*}
  \of{\dom}{B} &\subseteq \of{\dom}{A}&
  \sem{\tuple{P_\mn}}{\mdl B} &\subseteq \sem{\tuple{P_\mn}}{\mdl A}\\
\sem{\tuple{P_\rs}}{\mdl B} &= \sem{\tuple{P_\rs}}{\mdl A} \restriction \dom(\mdl B) &
  \sem{\tuple{P_\fx}}{\mdl B} &= \sem{\tuple{P_\fx}}{\mdl A}\\
  \sem{\tuple{f_\rs}}{\mdl B} &= \sem{\tuple{f_\rs}}{\mdl A} \restriction \dom(\mdl B)
\end{align*}
The $\circordeq$-minimal models of a formula $\phi$ can be described by a second-order formula:
\begin{equation*}
\begin{split}
  \mc{C}(\phi) \dfn \phi \wedge
  \forall D,\tuple{M},\tuple{V}. 
  &\big(\form{dom}(D) \wedge \tuple P_\fx \subseteq D
                    \wedge \tuple{M} \subseteq \tuple{P_\mn}
                  \wedge \phi[D]\{\tuple{M}/\tuple P_\mn, \tuple V / \tuple P_\vr \}\big)\\
                  &\rightarrow (\tuple{P_\mn} \subseteq \tuple{M} \wedge \forall x. D(x))
\end{split}
\end{equation*}
where $D$ is a unary predicate symbol and $(\tuple M, \tuple V)$ is similar to $(\tuple P_\mn, \tuple P_\vr)$\@. $\form{dom}(D)$ specifies that $D$ behaves like a domain 
(closed by function application, nonempty), 
$\phi[D]\{\tuple M / \tuple P_\mn, \tuple V / \tuple P_\vr\}$ is $\phi$ with all quantifications relativised by $D$ (e.g. $\fa{x} \psi$ becomes $\fa{x \in D} \psi$), and symbols in $\tuple M, \tuple V$ substituting symbols in $\tuple P_\mn, \tuple P_\vr$\@. $\tuple{P}_\fx \subseteq D$ means that every component of every relation in $\tuple P_\fx$ is in $D$, and for $\tuple A, \tuple B$ two similar relational tuples, $\tuple A \subseteq \tuple B$ is the componentwise inclusion.

\begin{theorem}\label{p:l:circum+ok}
  With $\phi$ a first-order formula on $\Upsilon$, the models of $\mc{C}(\phi)$ are the $\circordeq$-minimal models of $\phi$\@.
\end{theorem}
The proof builds upon \cite{Doherty1998}. Given a model $\mdl A$ of $\mc{C}(\phi)$ and $\mdl B \circordeq \mdl A$, the internal structure of $\mdl B$ can be ``plugged in'' the tuple $D,\tuple M, \tuple V$ and verifies
the left-hand side of the main implication in $\mc{C}(\phi)$; the right-hand side implies that $\mdl B$ cannot be strictly smaller than $\mdl A$\@. 
For the other direction, with $\mdl A$ a $\circordeq$-minimal model of $\phi$, we construct models from any $D,\tuple M,\tuple V$ that verify the antecedent, and by minimality of $\mdl A$ show that they verify the consequent. It is easy to see that $\tuple P_\mn$ can also contain FO formulas that use fixed or varying predicates \cite{Etherington1986}.

\subsection{Application of unified circumscription to transitions}\label{s:circum_application}

Let $\Theta \dfn (\tuple{\Dyn},\pt{\tuple{\Dyn}},\tuple{\Stat})$ be a transition signature. Let $\tuple{\Stat} = \tuple{g},\tuple{K}$ with $\tuple g$ purely functional and $\tuple K$ purely relational. Let $\Delta\tuple{\Dyn}$ be the tuple of formulas of the form $\Delta\form{A}(\tuple x)$ for $\form A \in \tuple{\Dyn}$\@.
Consider the circumscription order $\circordeq$ induced by the following mapping:
\begin{align*}
  \tuple{P_\mn} &\dfn \Delta \tuple{\Dyn} & 
  \tuple{P_\rs} &\dfn \tuple N & 
  \tuple{P_\vr} &\dfn \tuple{\pt{\tuple{\Dyn}}}&
  \tuple{P_\fx} &\dfn \tuple{\Stat} & 
  \tuple{f_\rs} &\dfn \tuple g & &
\end{align*}
That is, the precondition of a transition is fixed ($\tuple P_\fx$), static information on the remaining elements may not change ($\tuple N$, $\tuple g$), the postcondition can change freely ($\pt{\tuple{\Dyn}}$), and both the domain and changes are minimised ($\Delta\tuple{\Dyn}$).
We check that the change ordering is actually an instanciation of unified circumscription:
\begin{lemma}\label{p:l:orders+coincide}
For $\mdl A, \mdl B:\Theta$,
$\mdl B \circordeq \mdl A$ iff $\mdl B \actordeq \mdl A$\@.
\end{lemma}
The proof is a trivial unrolling of the definitions of $\circordeq$ and $\actordeq$\@. As an immediate corollary of theorem \ref{p:l:circum+ok} and lemma \ref{p:l:orders+coincide}, for any $\Theta$-formula $\phi$, $\mc{C}(\phi) \equiv\; \circum\phi$.

\subsection{Main theorem}\label{s:main+theorem}
\begin{lemma}\label{p:l:preserved+is+fo+expr}
  If $\phi$ is preserved under $\safe\consep d$ then $\circum(\phi\wedge\Tns)$ is first-order expressible.
\end{lemma}
The proof gradually removes second-order quantification from $\circum\phi$ (cf. section \ref{s:circumscription}). First, the restriction to FLBs (by $\Tn$) removes the universal quantification on $\tuple V$\@. Next, the domain is covered by $\form P$ and $\pt{\form{P}}$ (by $\textit{Support}$), so the universal quantification on $D$ can be removed. Next we show that minimal models of preserved formulas have changes localised around the images of the variables in $\safe$ and within a radius $d$\@. With this bound on the changes present in the minimal models of $\phi\wedge\Tns$, the universal quantification on $\tuple M$ can be replaced with first-order quantification. This translation is global and not compositional as in e.g. the reduction of some modal logics to FO.

\begin{lemma}\label{p:l:preserved+fragment}
  If $\phi$ is preserved under $\safe\consep d$, in $\exists^*\forall^*$ and $\forall^*\exists^*$, then $\circum(\phi\wedge\Tns)$ is in $\exists^*\forall^*$ and $\forall^*\exists^*$\@.
\end{lemma}
A refinement of lemma \ref{p:l:preserved+is+fo+expr}. The proof of this lemma exploits the locality of changes and the functionality of $\form{Link}$ and $\pt{\form{Link}}$ to switch quantifiers as necessary: modulo functionality of $R$, $\fa{y} R(x,y) \rightarrow \psi(x,y) \equiv \left(\fa{x} \neg R(x,u)\right) \vee \left(\ex{y} R(x,y) \wedge \psi(x,y)\right)$\@.

\vspace{1em}
\noindent\textit{Theorem \ref{l:main+theorem}}\ (restated)\\
  If $\safe\consep d\vdash\phi$ then $\phi \wedge \Tns \in \exists^*\forall^*$and $\phi\in\forall^*\exists^*$\@.

\vspace{1em}
\noindent Proof by induction on the derivation. The hard part is "rule:forall" when $\varof t = \varof u$; done by induction on the number of $\exists$ quantifiers below the new $\forall$\@. We use theorem \ref{l:proved+preserved} and lemma \ref{p:l:preserved+fragment} for "rule:circum". We again use functionality of $\form{Link}$ and $\pt{\form{Link}}$ for the other cases.

\section{Related work} 

Circumscription dates back to \cite{McCarthy1980}. We use an instanciation of unified circumscription, a new flavor of circumscription which generalises \cite{Doherty1998}. Previous works on taming circumscription require global syntactic properties of the formulas \cite{Doherty1998,nonnengart1999elimination,Conradie2006} and only consider satisfiability or FO-expressivity of circumscribed formulas.
\cite{Doherty2004} uses circumscription for characterising weakest preconditions to reactions. 
The Floyd-Hoare tradition extends to e.g. separation logic \cite{reynolds2002separation}, with an emphasis on model checking, and can allow more than 2 states, which can be first-class or modal \cite{Reiter1991,harel2001dynamic}, with a focus on program traces.

There are biological knowledge bases with different degrees of formalism \cite{uniprot,biopax}. Other modelling uses resource-aware logics \cite{Despeyroux2016,Boniolo2010}, or logic rules for specification and modality for queries \cite{pathwayLogic}. Full expressivity comparison with existing logics of changes (Hoare-like, modal, etc) would require more space than currently available.
%
\section{Conclusion and future work}\label{s:conclusion}
We have introduced a framework for describing and reasoning with molecular biology knowledge. 
We follow the tradition of taking graph rewriting as a domain-specific language for biology \cite{kappa,BNGL}.
Biological entities are described at the level of proteins in the form of bounded trees containing encodings of domains, subdomains and residues. Links between the trees represent protein-protein interactions. Proteins and their parts have both static and dynamic properties. Formulas represent observations of changes as a pair of forests $\langle$\textit{Precondition}, \textit{Postcondition}$\rangle$ with shared underlying sets. The theory of forest transitions is $\Tns$\@. This theory does not have decidable satisfiability, but modulo $\Tns$, the $\exists^*\forall^*$ fragment has.

As a knowledge representation tool, the logic describes changes in a compositional way, and a closed-world assumption on changes can be applied with a minimisation operator $\downarrow$, defined using a variant of circumscription. A proof system produces formulas that can be queried, in the sense that validity and satisfiability are decidable, including minimised formulas, which a priori were only second-order expressible.

The proof uses a semantic property, \emph{preservation}, to ensure that the change-minimal models of deducible formulas are first-order expressible. In addition, syntactic manipulation modulo $\Tns$ shows that deducible formulas are in the fragments $\exists^*\forall^*$ and $\forall^*\exists^*$\@. 

Importantly, some formulas with unguarded existential quantifiers can be first-order circumscribed. As future work we plan to extend the definition of preservation to capture a larger class of formulas.  
In ongoing work, we continue the development of this framework. In particular, we wish to identify a logical fragment where automatic synthesis of graph rewriting rules from $\circum$-minimised specifications becomes a possibility. The hope is to assist and partly automate biological modelling, from the description of observations at a high level of abstraction, to the execution of simulations and the validation of hypotheses. Future research also includes optimising the compilation to first-order 
and introducing reaction rates, i.e. transitions weights between the preconditions and postconditions.
\bibliographystyle{eptcs}
\bibliography{biblio}





\end{document}